\documentclass[conference,a4paper]{IEEEtran}
\IEEEoverridecommandlockouts
\usepackage{cite}
\usepackage{amsmath,amssymb,amsfonts}
\usepackage{algorithmic}
\usepackage{graphicx}
\usepackage{textcomp}
\usepackage{xcolor}
\usepackage{booktabs}
\usepackage{multirow}
\usepackage{array}
\usepackage{url}
\usepackage[colorlinks=true,linkcolor=blue,citecolor=blue,urlcolor=blue]{hyperref}

\def\BibTeX{{\rm B\kern-.05em{\sc i\kern-.025em b}\kern-.08em
    T\kern-.1667em\lower.7ex\hbox{E}\kern-.125emX}}

\begin{document}
\pagestyle{plain}

\title{Deep Learning-based Amharic Chatbot for FAQs in Universities}

\author{\IEEEauthorblockN{Goitom Ybrah Hailu}
\IEEEauthorblockA{\textit{Department of Computing} \\
\textit{Aksum University}\\
Aksum, Ethiopia \\
goitom50@gmail.com}
\and
\IEEEauthorblockN{Hadush Hailu Gebrerufael}
\IEEEauthorblockA{\textit{Department of Computer Science} \\
\textit{Tsukuba University}\\
Tsukuba, Japan \\
had.hailu@gmail.com}
\and
\IEEEauthorblockN{Shishay Welay (PhD.)}
\IEEEauthorblockA{\textit{School of Computing} \\
\textit{Mekelle University}\\
Mekelle, Ethiopia \\
bonjour.mit@gmail.com}
}

\maketitle \thispagestyle{plain}

\begin{abstract}
University students often spend a considerable amount of time seeking answers to common questions from administrators or teachers. This can become tedious for both parties, leading to a need for a solution. In response, this paper proposes a chatbot model that utilizes natural language processing and deep learning techniques to answer frequently asked questions (FAQs) in the Amharic language. Chatbots are computer programs that simulate human conversation through the use of artificial intelligence (AI), acting as a virtual assistant to handle questions and other tasks. The proposed chatbot program employs tokenization, normalization, stop word removal, and stemming to analyze and categorize Amharic input sentences. Three machine learning model algorithms were used to classify tokens and retrieve appropriate responses: Support Vector Machine (SVM), Multinomial Na\"{i}ve Bayes, and deep neural networks implemented through TensorFlow, Keras, and NLTK. The deep learning model achieved the best results with 91.55\% accuracy and a validation loss of 0.3548 using an Adam optimizer and SoftMax activation function. The chatbot model was integrated with Facebook Messenger and deployed on a Heroku server for 24-hour accessibility. The experimental results demonstrate that the chatbot framework achieved its objectives and effectively addressed challenges such as Amharic Fidel variation, morphological variation, and lexical gaps. Future research could explore the integration of Amharic WordNet to narrow the lexical gap and support more complex questions.
\end{abstract}


\section{Introduction}
Artificial Intelligence (AI) has been a hot topic since its inception in 1956. Its ultimate goal is to create intelligent machines that can think and act like humans. AI can be implemented in almost every sphere of work, and intelligent agents can do many tasks, from labor work to sophisticated operations \cite{ali2018multiagent, khanna2015study}. Natural Language Processing (NLP) is a field of study that focuses on the interaction between computers and humans in a natural way, which includes many natural language-related topics such as sentiment analysis, text similarity, question answering, and text summarization \cite{khurana2017nlp}.

One of the applications of AI that has gained momentum in recent years is chatbots. Chatbots seek to mimic human conversation, and their architecture combines a language model and machine learning algorithms to provide an informal communication channel between a human user and a machine. Chatbots are being used in many fields, from customer service and knowledge collection to entertainment, and they have the potential to enhance user experience and communication \cite{shawar2007chatbots}.

However, despite the popularity of chatbots, there is no research on Amharic chatbots. The Amharic language poses challenges due to its morphological richness and the shortage of resources. Therefore, this article proposes a study to develop an Amharic chatbot using deep learning techniques to facilitate instant information retrieval for users. The study aims to promote better interactivity, sociability, and knowledge acquisition in higher education, and to provide a wide range of services for mobile users who spend most of their time on email and messaging platforms such as Telegram and Messenger.

\subsection{Problem Background}
Nowadays, universities handle a large number of regular requests from students and others seeking information and answers to commonly asked questions. This often requires a dedicated support staff and can be time-consuming for students who need to physically visit various offices to seek information. In addition, administrators and teachers are burdened with the tiresome obligation of answering the same questions repeatedly and conducting numerous meetings with students to address their concerns. While email is an effective medium for providing information to a large number of students, it can be slow and ineffective for handling single requests or specific problems. Students also waste time searching for frequently asked questions on different websites and web pages.

To address these challenges, artificial intelligence in the form of chatbots can provide a solution. Chatbots can simulate a conversation with users in natural language and provide immediate and up-to-date information. However, developing chatbots for languages such as Amharic poses a challenge as the language has a different form of grammatical structure, representation of characters, and formation of statements. While there are chatbots developed for other languages, to the best of our knowledge, there is no Amharic chatbot for FAQs developed so far. Therefore, building a chatbot that can translate and understand multiple languages, including Amharic, requires additional time and effort during the designing and development phase. In addition, it is sometimes challenging for translator services to differentiate between languages with the same script, leading to difficulties in detecting the right language when users use both languages in a phrase. Additionally, chatbots must be aware of the end-user's culture, able to understand regional tones, and the ability to understand conversations and different regional accents or linguistic varieties (Dialects).

\subsection{Related Work}
In this section, we review the existing research on chatbots and question answering systems, with a focus on the Amharic language.

One notable related work is the FAQchat system, which retrained the ALICE system using the frequently asked questions (FAQs) from the School of Computing at the University of Leeds. FAQchat employed keyword-based retrieval without linguistic tools or meaning analysis. Users found it preferable over Google due to its direct answers and fewer links \cite{shawar2005faqchat}.

Another study described the design and development of a chatbot for university FAQs \cite{ranoliya2017chatbot}. It utilized Artificial Intelligence Markup Language (AIML) and Latent Semantic Analysis (LSA). AIML handled template-based and general questions, while LSA addressed service-based questions. The chatbot operated in three steps: user query processing, predefined format matching, and pattern-based answer presentation. However, this rule-based chatbot lacked the ability to learn new input data.

A conversational chatbot model was proposed as a substitution for industry FAQ pages \cite{nair2018faq}. It controlled the conversation flow based on user requests and provided natural language responses, including direct answers, requests for additional information, or recommended actions. The model employed deep learning techniques for intent and entity recognition.

In another study, a framework was developed for a chatbot that could be used by independent companies as a customer support replacement \cite{singh2018chatbot}. The framework incorporated AI at its core, utilizing TensorFlow to create a neural network trained with a design document for response generation. The system consisted of a user interface, neural network model, NLP unit, and feedback system.

Researchers have also explored Arabic chatbots that employ AIML pattern matching for FAQ answering \cite{shawar2009arabic}. While achieving a high correctness rate for Arabic questions, these systems faced challenges when dealing with different Arabic forms.

The first Arabic chatbot, BOTTA, was developed using AIML on the Pandorabots platform \cite{ali2016botta}. BOTTA aimed to simulate conversation and engage with Arabic language users.

Additionally, there have been efforts to develop Amharic factoid and non-factoid question answering systems \cite{yimam2009teteyeq, zeleke2013leteyeq, abedissa2013amharic}. These systems employed various techniques such as preprocessing, question analysis, document retrieval, and response extraction. However, no comprehensive chatbot capable of answering Amharic FAQs for students has been developed.

In summary, the related works have explored different aspects of chatbot and question answering systems. However, there is still a need for a robust chatbot capable of addressing FAQs in the Amharic language. In comparison to existing chatbot systems, the proposed Amharic chatbot offers several distinct advantages. While previous systems relied on keyword-based retrieval and lacked linguistic tools or meaning analysis, our chatbot leverages deep learning techniques to provide more accurate and contextually meaningful responses. This study aims to bridge this gap by utilizing deep learning techniques to develop a comprehensive chatbot solution that supports Amharic language. Moreover, unlike other chatbots that are primarily designed for widely spoken languages, our Amharic chatbot addresses the specific challenges of an under-resourced language, making it a valuable tool for the Amharic-speaking community. The user feedback and comparative evaluations have shown that our Amharic chatbot outperforms general-purpose chatbots in terms of providing direct answers, reducing reliance on external links, and delivering a more personalized user experience.

\section{Methodology}
The research methodology used in this study was Design Science Research (DSR) \cite{hevner2004design}, which focuses on creating artifacts that can solve practical business problems. DSR aims to generate scientific knowledge while developing technology-based solutions. It consists of three cycles: the Relevance Cycle, the Design Science Research Cycle, and the Rigor Cycle. The Relevance Cycle connects the research context with the design science activities, defining the problem space and acceptance criteria. The Design Science Research Cycle involves building and evaluating artifacts using computational and mathematical methods. The Rigor Cycle connects design science activities with existing knowledge and foundations. The research also follows a six-step design science research methodology process model for information systems formulated by Peffers et al.\ \cite{peffers2007design}. These six-step process model are: Problem Identification and Motivation, Objectives of the Study, Design and Development, Demonstration, Evaluation, and Communication. The problem is identified as the lack of an Amharic chatbot for FAQs in universities. The objective is to design and implement a deep learning-based Amharic chatbot. The design and development activity focus on creating the chatbot framework using different tools and techniques.

\subsection{Data Collection}
To develop a domain-specific chatbot, a dataset is required to test and validate the chatbot's performance. The researcher chose engineering students as the target audience because they have multiple questions related to their department, selecting engineering departments, and other related questions that ensure the breadth of the dataset.

A purposive sampling method was used to collect the dataset for the chatbot model. Purposive sampling is a sampling technique that relies on the researcher's judgment when selecting the units to be studied. A questionnaire was distributed to engineering students in Mekelle University and Aksum University. The questionnaire was completed by 80 students, 38 males and 12 females from Mekelle University and 13 males and 17 females from Aksum University.

The questionnaire aimed to identify frequently asked questions by students in engineering departments during their studies. The collected data was used as a basis for determining the Amharic training data. The collected data was translated, preprocessed, and corrected to Amharic language using Google Translator and some Amharic language experts.

After collecting the questionnaires, the dataset information required for the chatbot was extracted, and the dataset was narrowed down to cover 60 topics. The collected datasets were structured into a JSON file. JavaScript Object Notation (JSON) uses human-readable text to store and transmit data objects consisting of attribute-value pairs and array data types. With the JSON package in Python, the JSON file was able to be read and be prepared to be processed and used for training. The series of intents consist of tags, patterns, responses, context set and context filter.

Each intent entry consists of a \textit{tag} (a unique name for each of the 60 topics), \textit{patterns} (sample queries for each topic), \textit{responses} (candidate answers from which one is randomly selected after identifying the topic), \textit{context set} (which changes the conversation state if needed), and \textit{context filter} (which filters results based on the current context).

This JSON structure model chat data has two main advantages. First, if someone has a chat dataset, every query can be marked for a tag. The training dataset can thus be attached to existing query data and answers. Secondly, if there is no previous dataset, anyone can be able to create and add data in that format. The conversation's flow is no problem to determine. If anyone wants to mark a dialogue about other data, the `Context filter' tag will be followed.

\subsection{Chatbot Framework Design}
\begin{figure}[htbp]
\centerline{\includegraphics[width=\columnwidth]{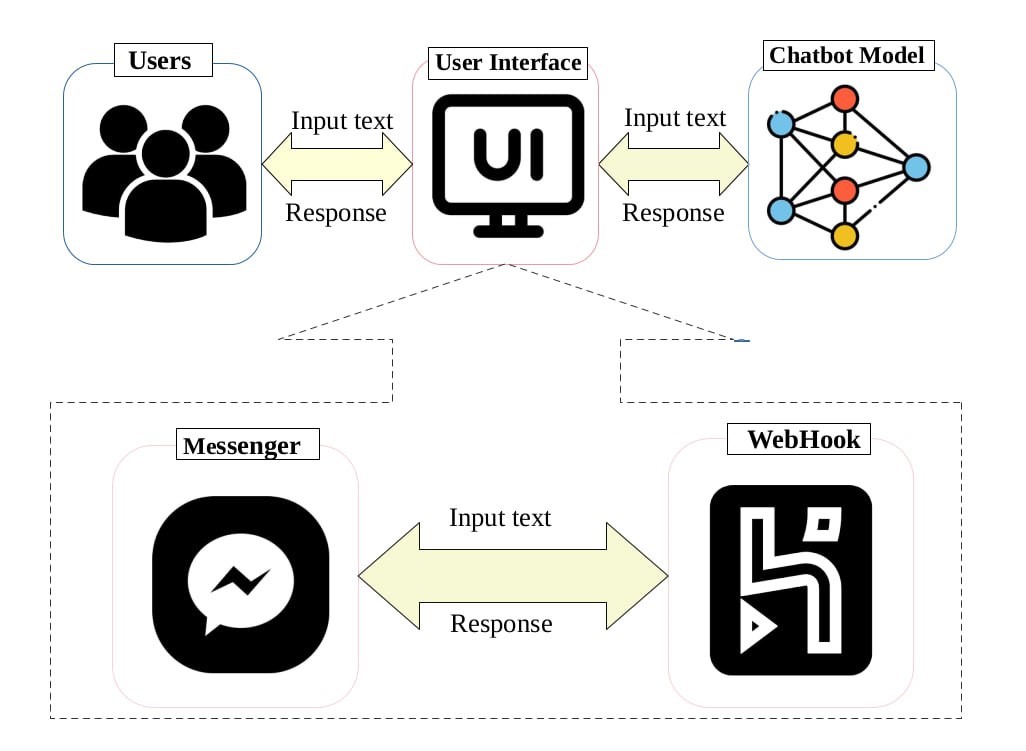}}
\caption{Conceptual framework of FAQ chatbot.}
\label{fig:conceptual}
\end{figure}

The conceptual chatbot framework consists of three main parts: the user, the user interface, and the chatbot model as shown in Fig.~\ref{fig:conceptual}.

The user part defines the end-users or students who need answers and information about their studying and other related questions in the university. The user interface (UI) part describes how a user communicates and interacts with the chatbot. It is a series of elements of natural languages that allow for interaction between user-chatbot models. This means users can communicate on their terms, not the computers. However, a chatbot's communication skill can vary depending on the interface created. A chatbot interface that uses default answers, like button options, limits the question the user can ask and understands the chatbot. But the chatbot that is built in this paper was designed to understand and respond to a variety of Amharic text inputs through a Facebook Messenger which acts as an interface and interacts to the chatbot model using a Flask webhook; and deployed it in Heroku web server to have a real-time conversation.

\begin{figure}[htbp]
\centerline{\includegraphics[width=\columnwidth]{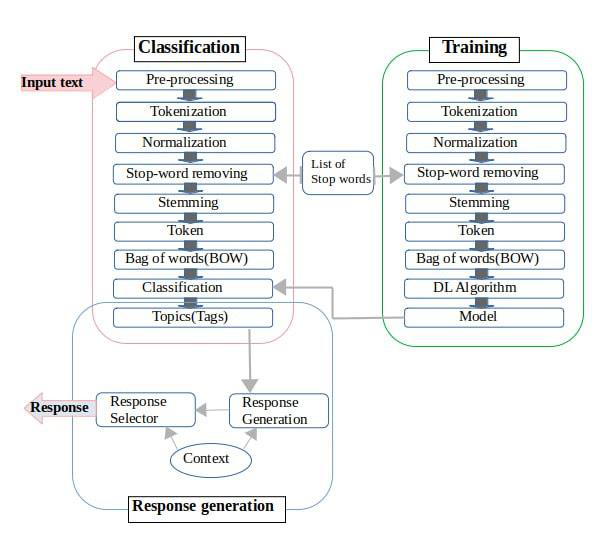}}
\caption{Amharic chatbot model.}
\label{fig:chatbot_model}
\end{figure}

The chatbot model part designed in this paper, as shown in Fig.~\ref{fig:chatbot_model}, consisted of three main parts: intent classification, training, and response generation.

In the classification part, preprocessing techniques such as tokenization, normalization, stop word removal, and stemming were applied to the user questions. This preprocessing helped in identifying the intent of the questions and retrieving appropriate responses. The dataset underwent preprocessing and stemming, and a classification model was trained to assign class labels to new queries and provide suitable responses.

The training part involved preprocessing and feeding the collected dataset to the classification model for training and feature extraction.

The response generation part focused on providing appropriate responses based on the data provided in the classification part. The response generation part involves selecting an appropriate response based on the identified intent. This can be done by selecting a random response from the list of responses for the identified intent or by using a more advanced response generation algorithm, such as a sequence-to-sequence model, to generate a more natural and contextual response.

The overall design of the chatbot framework model is focused on the chatbot text classification model, and the researcher discusses and tests different text classification methods and algorithms such as SVM, MNB and DNN to learn the effectiveness of different algorithms for finding out the best possible text classifier for the Amharic dataset, and the researcher compared and chose the best performing one which is the DNN as the final intent classifier for the chatbot classification model. The deep neural network (DNN) we used is the Sequential deep learning model from TensorFlow and Keras, the Python Deep Learning Library. This type of model is a supervised learning model. To create the Sequential model, we used the adding method to add 4 dense layers that are densely connected. The first input layer has neurons equal to the number of features (vectors) used to train the classifier, the two dense hidden layers have 128 and 64 hidden neurons, and the last layer is an output layer equal to the number of intents to predict output intent. We use ReLU as the activation function for the two hidden layers and softmax for the fourth layer which is the output layer.

To improve the model prediction and reduce overfitting, the researcher applied dropout layers to the model with the probability of retaining the unit being 0.2 for the hidden layers and a learning rate of 0.0001 to the Adam optimizer which produced a good result for the model. This made the deep neural network better at generalization and less likely to overfit the training data.

To compile the model, the researchers also used the Adam optimizer \cite{kingma2015adam} to train the neural networks as it has been shown to produce good results in the field of deep learning. It is a specialized gradient-descent algorithm that uses the computed gradient, its statistics, and its historical values to take small steps in its opposite direction inside the input parameter space as a means of minimizing a function. It is also computationally efficient which is particularly useful here due to the number of samples to optimize. To fit and save the model, we used the fit function. The model is trained for 150 epochs to achieve good accuracy and batch size of 8.

\begin{figure}[htbp]
\centerline{\includegraphics[width=\columnwidth]{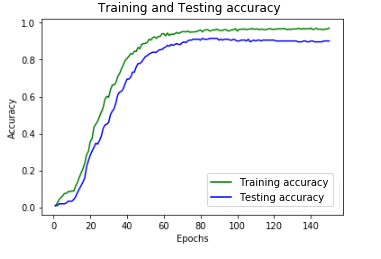}}
\caption{Training and testing accuracy and loss output.}
\label{fig:training}
\end{figure}

Fig.~\ref{fig:training} shows the training output of the model for 150 epochs with the model loss.

After the model is built and trained, the next step is to predict a response for the user. To do this, the user input has to be processed first. The method we use to process the user input is similar to how we processed the training data for the model. We took in the user input and used the NLTK tokenizer to create a vector of words. Then, we used the Bag of Words method for similarity measure so we can easily compare the user input to the questions in our model. We use the Keras predict function in the model class API to calculate the probability of the user inputted question being in each class in our database. Then we find the class with greatest probability and return one of the responses to the user. During the compiling of the model, it outputs information from each epoch. The loss, accuracy and mean squared error for each epoch is outputted. Lastly, the chatbot model built is integrated with Facebook Messenger and deployed it in the Heroku server to be accessible and active 24 hrs for the students.

\section{Experiments and Results}
This research paper focused on building a chatbot that can understand the Amharic language and provide answers to the most frequently asked questions by students in the engineering departments of universities. The study collected data by distributing a questionnaire to students to gather frequently asked questions. The study aimed to provide high-quality answers to these questions, and it built and compared three different chatbot models to achieve this objective.

\begin{figure*}[htbp]
\centerline{\includegraphics[width=0.75\textwidth]{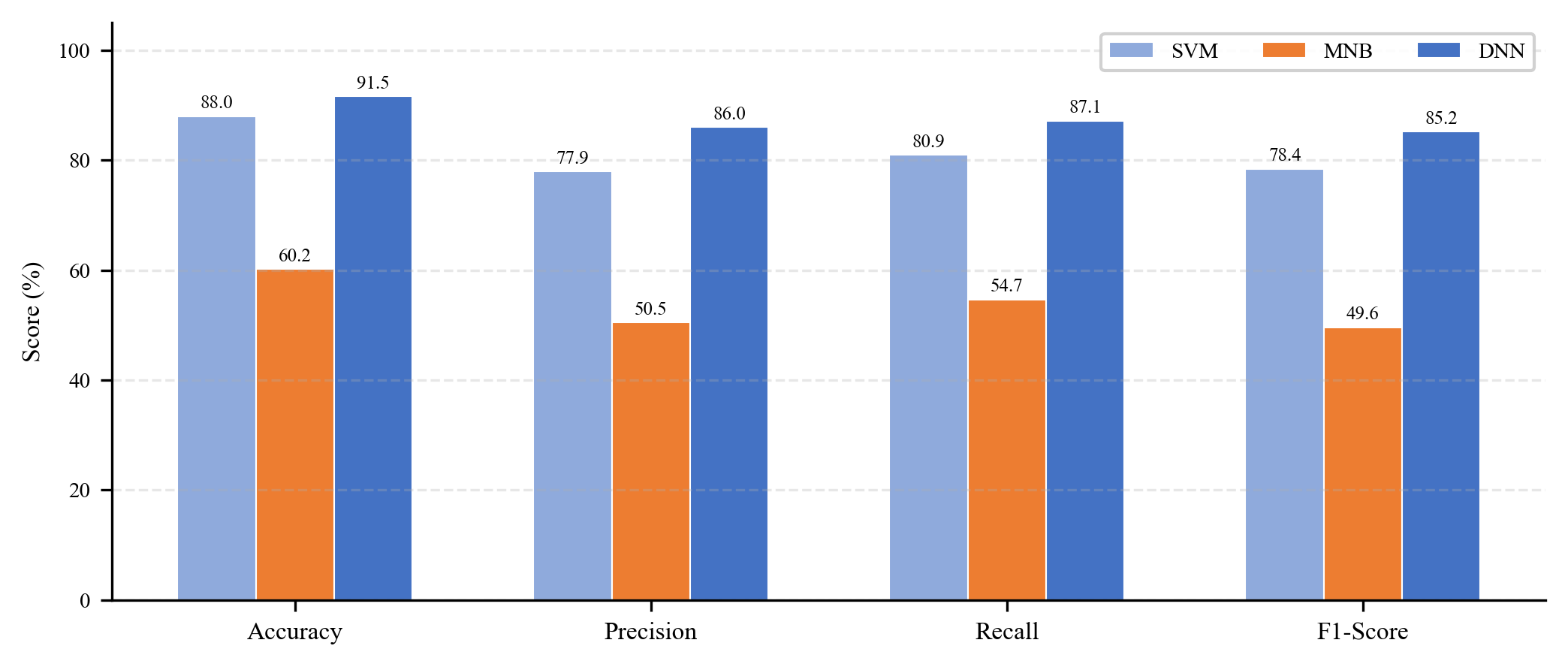}}
\caption{Text classification algorithms evaluation.}
\label{fig:classifiers}
\end{figure*}

The best performing chatbot model had a 91.55\% accuracy rate and an 85.21\% F1-score, with a loss value of 0.3548. The experimental results demonstrated the effectiveness of the proposed chatbot system.

To evaluate the performance of the chatbot model, the researchers used various evaluation metrics, such as accuracy, precision, recall, and F1 score, which are renowned for measuring the performance of classification and information retrieval tasks. The precision, recall, and F1 score analysis revealed the chatbot's robustness in accurately categorizing and retrieving appropriate responses to Amharic user queries.

The researchers used a machine learning approach to create the chatbot. While some natural language processing (NLP) functions were used, the response generation process relied heavily on machine learning, as chatbots based on machine learning do not understand the meaning of sentences; they learn how to respond based on previous experience. The performance of the intent classification had an effect on the response generation part, and the overall performance of the intent classifier was measured by computing the evaluation metrics of a particular classifier on a testing dataset.

The researchers prepared 850 pattern queries for 60 tags (intents) from the students via questionnaire, and they used 20\% of the data for testing. To evaluate the performance of the three classifiers (SVM, MNB, and DNN), the researchers used basic evaluation metrics such as accuracy, precision, recall, and F1-score. The experimental evaluation showed that the DNN classifier performed better than the other types of classifiers, with a 91.55\% accuracy rate, 85.98\% precision, 87.13\% recall, and 85.23\% F1-score.

The researchers also tested the performance of the SVM and MNB algorithms, and all results are shown in Fig.~\ref{fig:classifiers}.

The SVM algorithm had an accuracy rate of 87.96\%, while MNB had an accuracy rate of 60.19\%. As the researchers did not know which algorithm was the best for text classification, they tested various activation functions on the DNN model to find out which one had the best performance. The results are shown in Table~\ref{tab:activations}, and the Rectified Linear Unit (ReLU) had the best performance, with a 91.55\% accuracy rate, 85.98\% precision, 87.13\% recall, and 85.23\% F1-score.

\begin{table}[htbp]
\caption{Activation Functions}
\label{tab:activations}
\centering
\footnotesize
\begin{tabular}{p{3cm}cccc}
\toprule
Activation functions applied to the DNN model & Accuracy & Precision & Recall & F1-Score \\
\midrule
ReLU    & 91.55\% & 85.98\% & 87.13\% & 85.23\% \\
Sigmoid & 69.95\% & 65.00\% & 66.36\% & 62.89\% \\
Tanh    & 84.98\% & 82.45\% & 81.14\% & 80.04\% \\
Linear  & 87.79\% & 83.25\% & 82.13\% & 81.16\% \\
\bottomrule
\end{tabular}
\end{table}

After conducting several experiments by changing the features of the DNN model, the researchers found that the model with two hidden layers, which had 128 and 64 nodes, respectively, gave the best testing results for 150 epochs.

\begin{figure*}[htbp]
\centerline{\includegraphics[width=\textwidth]{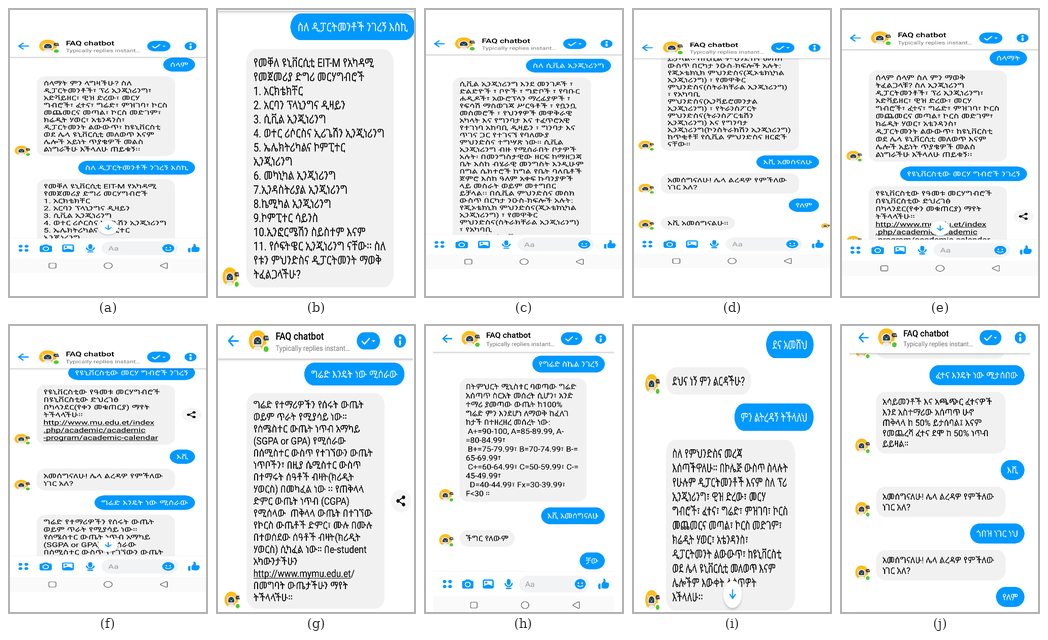}}
\caption{Prototype sample: sequential conversation screenshots demonstrating the Amharic chatbot interaction via Facebook Messenger.}
\label{fig:prototype}
\end{figure*}

In conclusion, the study developed a chatbot model that can understand the Amharic language and provide high-quality answers to frequently asked questions in engineering departments. The model used machine learning techniques, with a focus on intent classification and natural language processing. The researchers compared three different chatbot models and found that the DNN model with the Rectified Linear Unit (ReLU) activation function and two hidden layers was the best performing model, with a 91.55\% accuracy rate. The study's results demonstrate the potential for chatbots to provide useful support for students in their studies. Fig.~\ref{fig:prototype} shows the GUI and examples of conversation threads.

\subsection{Evaluation}
The study conducted an evaluation of the developed chatbot model for providing information related to higher education studies in Ethiopia. Since all students of higher education had left the university due to the COVID-19 case, the evaluation was conducted randomly in a social network with 15 people equally distributed over the range of fresh and senior Engineering students and lecturers through a Google Form. The participants were composed of 11 males and 4 females with ages ranging from 20 to 32 years; four from the fresh engineering department, two from the second year Electrical department, two from the fourth year Civil department, four from the Computing department (Computer Science and IT) and three university lecturers. The participants were provided with information about the knowledge base of the chatbot and were asked to interact with it by asking questions related to their domain in Amharic language.

The study utilized the Likert scale to measure the satisfaction of the framework, where respondents were requested to show the level of their agreement or disagreement with the statements on a 5-point scale. The results of the evaluation revealed that the chatbot performed efficiently when the queries were simple and under the intent of the file created. Moreover, the prototype was evaluated based on various dimensions such as Input question grammar restriction (Q1), Robustness (Q2), Amharic alphabet variation support (Q3), Ease of Use (Q4), User friendliness (Q5) and Answer validity (Q6).

The user satisfaction score obtained based on the satisfaction measure questionnaire formulated revealed that the overall user satisfaction level was 86.2\%, indicating that the chatbot was successful in providing users with the required information and answering their queries related to higher education studies in Ethiopia as shown in Table~\ref{tab:satisfaction}. The users found the chatbot model to be flexible, user-friendly, easy to understand, and effective in providing accurate answers.

\begin{table*}[htbp]
\caption{The User Satisfaction Score Obtained Based on the Satisfaction Measure Questionnaire Formulated}
\label{tab:satisfaction}
\centering
\footnotesize
\renewcommand{\arraystretch}{1.2}
\begin{tabular}{lc|ccccc|cc}
\hline
 & & \multicolumn{5}{c|}{Scale Value} & & \\
\cline{3-7}
 & & 5 & 4 & 3 & 2 & 1 & & \\
\cline{3-7}
Questions & Total Respondents & Strongly Agree & Agree & Neutral & Disagree & Strongly Disagree & Average Values & Average in \% \\
\hline
Q1 & 15 & 0  & 8  & 6 & 1 & 0 & 3.47 & 69.4\% \\
Q2 & 15 & 1  & 13 & 1 & 0 & 0 & 4.00 & 80.0\% \\
Q3 & 15 & 12 & 3  & 0 & 0 & 0 & 4.80 & 96.0\% \\
Q4 & 15 & 14 & 1  & 0 & 0 & 0 & 4.93 & 98.6\% \\
Q5 & 15 & 12 & 3  & 0 & 0 & 0 & 4.80 & 96.0\% \\
Q6 & 15 & 2  & 9  & 4 & 0 & 0 & 3.87 & 77.4\% \\
\midrule
\multicolumn{7}{r|}{Total Average Value} & 4.31 & 86.2\% \\
\bottomrule
\end{tabular}
\end{table*}

In conclusion, the study demonstrated that the developed chatbot model was effective in providing users with the required information related to higher education studies in Ethiopia. The results of the evaluation indicated that users were highly satisfied with the chatbot model's performance, which could be further improved by incorporating additional features to enhance its performance.

\section{Conclusion}
This paper has successfully developed an Amharic chatbot framework that can understand and respond to frequently asked questions in the engineering department domain. The study has used deep learning techniques such as TensorFlow, Keras, and NLTK to design and develop the chatbot. The main contribution of this research work is the development of a framework that can be used to train and deploy chatbots with existing datasets easily. The study has also identified key components of Amharic chatbot systems and basic challenges that need to be addressed for the development of complete conversational Amharic chatbot systems.

The developed Amharic chatbot framework provides a significant contribution to the NLP field, as it enables target users to interact with the chatbot using their natural language. It also provides a significant contribution to Amharic language technology, as it is the first Amharic chatbot that can answer frequently asked questions in the university domain. The study has also identified basic challenges that need to be addressed to improve the chatbot's accuracy, such as the lack of large Amharic FAQ datasets, synonyms words, and the absence of Amharic grammar checker and spelling checker.

Overall, this research work can be considered as a starting point in the development of complete conversational Amharic chatbot systems. By integrating the findings of this study with future research works, the chatbot system can be further improved and developed into a complete conversational Amharic chatbot system that can understand and respond to various types of queries. Therefore, the developed Amharic chatbot framework can be used as input to develop complete multi-domain conversational Amharic chatbot systems.

\section{Contribution and Future Work}
The development of an Amharic chatbot for university students is a significant contribution to the field of conversational agents. The study successfully designed and compared three classifier algorithms for intent classification and showed how information extraction can be achieved in Amharic using standard deep learning and machine learning techniques. The study also identified the key components of Amharic chatbot systems, which can serve as a framework for developing FAQs in the university domain.

In future research, we aim to integrate sentiment analysis into the chatbot system to enhance its understanding of user emotions and provide more empathetic responses. Additionally, we plan to explore the possibility of expanding the language support of the chatbot to include other Ethiopian languages, which would further benefit the diverse user base in the region. Furthermore, scaling up the chatbot system to accommodate a larger user base and integrating voice recognition technology could significantly enhance the accessibility and usability of the chatbot, enabling users to interact with it through voice commands.

One of the main recommendations from the study is to extend the scope of the research to support complex questions and other question types, such as comparative or procedure questions, to make the chatbot smarter. Another recommendation is to use feedback neural networks like recurrent networks to improve the performance of the model by increasing the Amharic FAQs datasets. The study also recommends implementing the chatbot for various domains and adding sentiment analysis to the model to evaluate its performance in real-world scenarios.

Additionally, the chatbot can be improved for answering user-specific questions by using conditional design for some parts of the bot, such as increasing the tags of the dataset. Finally, enhancing the NLP techniques for the Amharic language and increasing the datasets can further improve the performance of the chatbot.

Overall, this study is an essential step forward in producing a complete conversational Amharic chatbot, and it can serve as a useful input for developing a multi-domain conversational Amharic chatbot by integrating it with this study. The recommendations provided in this study can guide future research on developing chatbots and QA systems for various natural languages, including Amharic.

\section*{Data Availability}
The dataset, trained model, and source code for this study are publicly available on GitHub at this \href{https://github.com/GoitomYbrah/Deep-Learning-Based-Amharic-Chatbot-for-FAQs-in-Universities-}{link}.

\bibliographystyle{IEEEtran}
\bibliography{references}

@article{ali2018multiagent,
  author    = {A. Ali and Z. Memon and A. H. Jalbani and M. Shaikh},
  title     = {Multi-Agent Communication System with Chatbots},
  journal   = {Mehran University Research Journal of Engineering and Technology},
  volume    = {37},
  number    = {3},
  pages     = {663--672},
  year      = {2018}
}

@article{khanna2015study,
  author    = {A. Khanna and B. Pandey and K. Vashishta and K. Kalia and B. Pradeepkumar and T. Das},
  title     = {A Study of Today's {A.I.} through Chatbots and Rediscovery of Machine Intelligence},
  journal   = {International Journal of u- and e-Service, Science and Technology},
  volume    = {8},
  number    = {7},
  pages     = {277--284},
  year      = {2015}
}

@article{khurana2017nlp,
  author    = {D. Khurana and A. Koli and K. Khatter and S. Singh},
  title     = {Natural Language Processing: State of The Art, Current Trends and Challenges},
  journal   = {Manav Rachna International University},
  year      = {2017}
}

@article{shawar2007chatbots,
  author    = {B. A. Shawar and E. Atwell},
  title     = {Chatbots: Are they Really Useful?},
  journal   = {ResearchGate},
  year      = {2007}
}

@inproceedings{shawar2005faqchat,
  author    = {B. A. Shawar and E. Atwell and A. Roberts},
  title     = {{FAQchat} as an Information Retrieval System},
  booktitle = {Proceedings of the Human Language Technologies Conference},
  year      = {2005}
}

@inproceedings{ranoliya2017chatbot,
  author    = {B. R. Ranoliya and N. Raghuwanshi and S. Singh},
  title     = {Chatbot for University Related {FAQs}},
  booktitle = {Proceedings of the IEEE International Conference on Advances in Computing, Communications and Informatics},
  pages     = {1525--1530},
  year      = {2017}
}

@article{nair2018faq,
  author    = {S. Nair and S. AD and S. SP and T. Sinha},
  title     = {Implementation of {FAQ} Pages using Chatbot},
  journal   = {International Journal of Computer Science and Information Security},
  volume    = {16},
  number    = {6},
  pages     = {187--194},
  year      = {2018}
}

@inproceedings{singh2018chatbot,
  author    = {R. Singh and M. Paste and N. Shinde and H. Patel and N. Mishra},
  title     = {Chatbot using {TensorFlow} for small Businesses},
  booktitle = {Second International Conference on Inventive Communication and Computational Technologies (ICICCT)},
  pages     = {1614--1619},
  year      = {2018}
}

@inproceedings{shawar2009arabic,
  author    = {Bayan Abu Shawar and Eric Atwell},
  title     = {Arabic question-answering via instance based learning from an {FAQ} corpus},
  booktitle = {CL2009 International Conference on Corpus Linguistics},
  year      = {2009}
}

@inproceedings{ali2016botta,
  author    = {D. A. Ali and N. Habash},
  title     = {{Botta}: An {Arabic} Dialect Chatbot},
  booktitle = {International Conference on Computational Linguistics},
  pages     = {208--212},
  year      = {2016}
}

@mastersthesis{yimam2009teteyeq,
  author    = {S. M. Yimam},
  title     = {{TETEYEQ}: {Amharic} Question Answering System for Factoid Questions},
  school    = {Addis Ababa University},
  year      = {2009}
}

@mastersthesis{zeleke2013leteyeq,
  author    = {D. Abebaw Zeleke},
  title     = {{LETEYEQ}: A Web Based {Amharic} Question Answering System for Factoid Questions Using Machine Learning Approach},
  school    = {Addis Ababa University},
  year      = {2013}
}

@mastersthesis{abedissa2013amharic,
  author    = {T. Abedissa},
  title     = {{Amharic} Question Answering For Definitional, Biographical and Description Questions},
  school    = {Addis Ababa University},
  year      = {2013}
}

@article{hevner2004design,
  author    = {A. R. Hevner and S. Ram and S. T. March and J. Park},
  title     = {Design Science in Information Systems Research},
  journal   = {MIS Quarterly},
  volume    = {10},
  number    = {2},
  pages     = {199--217},
  year      = {2004}
}

@article{peffers2007design,
  author    = {K. Peffers and T. Tuure and C. Samir},
  title     = {A Design Science Research Methodology for Information Systems Research},
  journal   = {Journal of Management Information Systems},
  volume    = {24},
  number    = {3},
  pages     = {45--78},
  year      = {2007}
}

@inproceedings{kingma2015adam,
  author    = {D. P. Kingma and J. L. Ba},
  title     = {{ADAM}: A Method for Stochastic Optimization},
  booktitle = {International Conference on Learning Representations (ICLR)},
  pages     = {1--15},
  year      = {2015}
}

\end{document}